\documentclass[twocolumn,english]{revtex4-1}
\usepackage[T1]{fontenc}
\usepackage[utf8]{inputenc}
\usepackage{color}
\usepackage{amstext}
\usepackage{graphicx}
\usepackage{wasysym}
\usepackage{babel}
\begin{document}

\title{Self-assembly of active core corona particles into highly ordered
and self-healing structures}

\author{Yunfei Du, Huijun Jiang, and Zhonghuai Hou{*}}

\address{Hefei National Laboratory for Physical Sciences at Microscales and
Department of Chemical Physics, iChEM, University of Science and Technology
of China, Hefei, Anhui 230026, China}

\email{hzhlj@ustc.edu.cn}

\begin{abstract}
Formation of highly ordered structures usually needs to overcome a
high free-energy barrier that is greatly beyond the ability of thermodynamic
fluctuation, such that the system would be easily trapped into a state
with many defects and the annealing process of which often occurs
on unreachable long time-scales. Here we report a fascinating example
theoretically that active core corona particles can successfully self-assemble
into a large-scaled and highly ordered stripe or trimer lattice, which
is hardly achieved in a non-driven equilibrium system. Besides, such
an activity-induced ordered structure shows an interesting self-healing
feature of defects. In addition, there exists an optimal level of
activity that most favorably enhance the formation of ordered self-assembly
structures. Since core corona particles act as important units for
self-assembly in real practice, we believe our study opens a new design-strategy
for highly ordered materials.
\end{abstract}
\maketitle

\section{Introduction }

Significant advancement has been made in self-assembly of colloidal
particles in recent decades\cite{boles2016self,batista2015nonadditivity}.
It is now possible to manipulate various types of colloidal particles
to fabricate them into highly ordered functional structures\cite{boles2015many,ye2015structural},
widely exploited in various fields such as in photonic\cite{hou2018patterned,sun2018design},
phononic\cite{li2018weyl,he2018topological} or lithographic applications\cite{rey2015fully}.
A popular choice recently for self-assembly unit is the so-called
core corona particle, which consists of an inside core (usually metallic
nanoparticle) surrounded by ambient soft corona (e.g., polymer chains
or microgels) to prevent the aggregation of cores\cite{si2018nanoparticle}.
The interaction between two core corona particles is characterized
by two repulsive length scales, related to the hard and soft repulsion,
respectively. It is reported that such interaction can result in phases
with non-trivial symmetries which include loose- and close-packed
hexagonal lattice, monomer, dimer, and trimer fluids, stripe and labyrinthine
phases, honeycomb lattice, etc\cite{malescio2003stripe,jagla1998phase}
and even quasicrystals with unexpectedly high symmetry\cite{dotera2014mosaic}.

Note that the highly ordered equilibrium phase of the system mentioned
above correspond to the state with lowest free energy that is thermodynamically
most stable for given parameters. Such ordered structures may have
great importance in material science, e.g., being used as templates
for lithography nanomanufacturing which requires zero-defect patterns
with even molecular level tolerance\cite{ruiz2008density}. In a real
process of self-assembly in experiment, however, one generally starts
from random initial conditions and the system would easily get trapped
into a metastable state that is not fully ordered but with many defects.
There typically exists a high free energy barrier between this metastable
state and the equilibrium ordered one, which is hardly overcome by
thermodynamic fluctuations\cite{rey2017anisotropic}, such that the
system would stay at the metastable state with defects for prohibitively
long time-scales\cite{bodnarchuk2011structural,aubret2018targeted}.
Therefore, how to obtain a highly ordered structure with low fraction
of defects as much as possible becomes a challenging problem in the
study of self-assembly.

Recently, dynamics of self-propelled active colloids which absorb
energy from ambient environment and push the system to a state far
from equilibrium, has gained extensive research attentions. A wealth
of remarkable collective behaviors have been reported both experimentally
and theoretically, including phase separation\cite{fily2012athermal,buttinoni2013dynamical,palacci2013living},
active turbulence\cite{dunkel2013fluid}, active swarming\cite{cohen2014emergent},
etc. Very recently, a few important works have noticed the perspective
of using active particle to alter the properties of colloidal aggregates\cite{mallory2018active}.
For example, A. Cacciuto $et.al$ used a collection of triangular
colloidal blocks with activity to achieve the self-assembly of capsid-like
structures\cite{mallory2016activity}. In a subsequent work, they
found that activity can also remarkably enhance the self-assembly
process of triblock Janus colloids into kagome lattice\cite{mallory2019activity}.
These works suggested that it is possible to design highly ordered
structures by introducing activity into the assembly unit.

\begin{figure*}
\begin{centering}
\includegraphics[width=1.7\columnwidth]{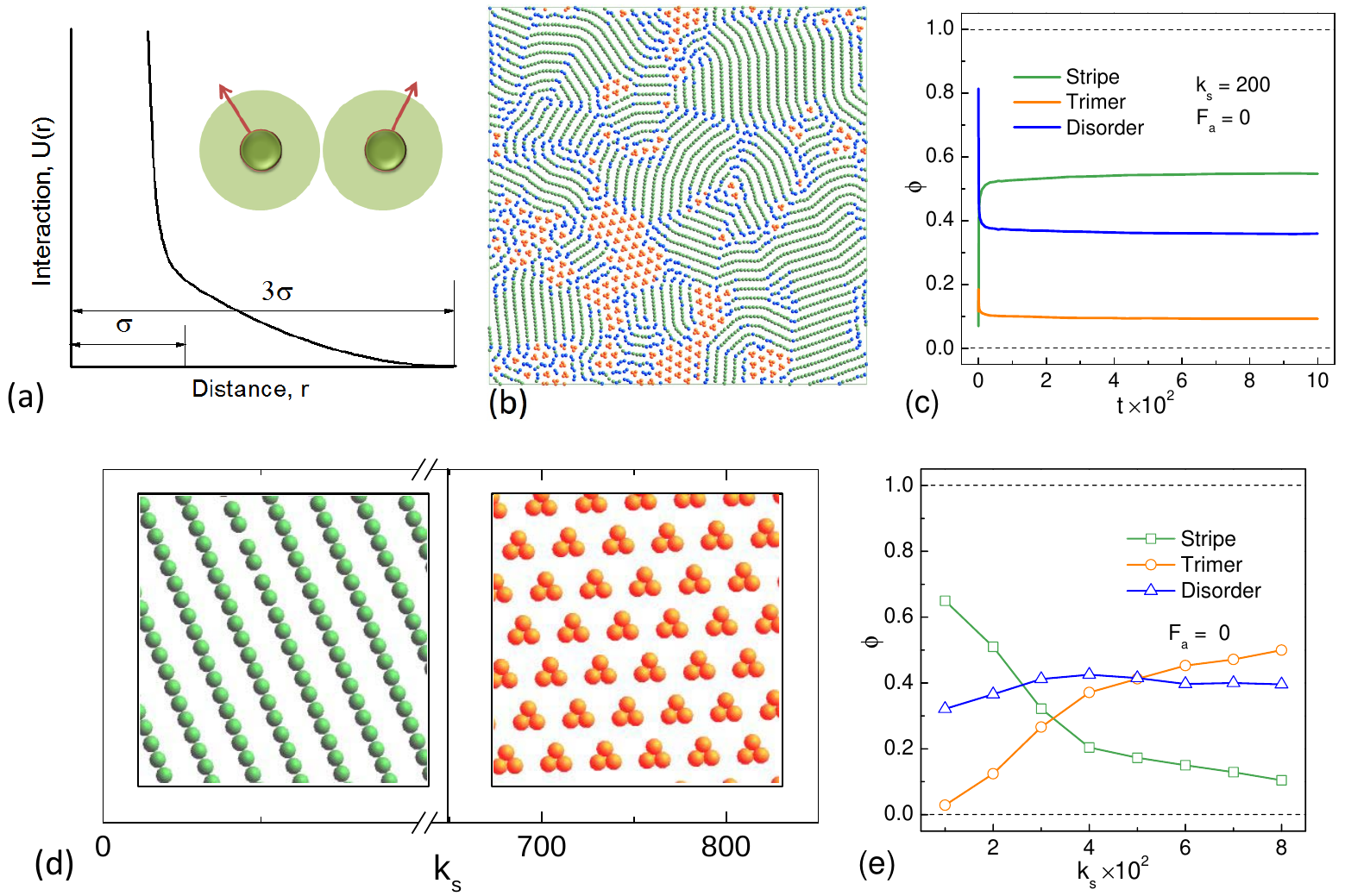}
\par\end{centering}

\caption{Model setting and the characteristics of a passive system. (a) Equivalent
interaction of a pair of active core corona particles. The core colored
in dark green is modeled as unpenetrable colloid which is propelled
by an active force $F_{a}$ depicted by red arrow. Note $F_{a}=0$
for (b)-(e), corresponding a passive case. The interaction between
the ambient coronas colored in light green is modeled by a simple
spring, which the strength of it is $k_{s}$. (b) Stable spatial configuration
of the system at $k_{s}=200$. Note that the coronas are not drawn
out for simplicity. Such a configuration consists of three typical
phases, namely stripe phase colored in green, trimer phase colored
in yellow and disorder phase colored in blue. (c) Fractions of three
phases $\phi_{i}(i=S,\,T,\,D$ corresponding to stripe, trimer and
disorder, respectively) over time. (d) Equilibrium phase diagram.
The phase boundary between stripe phase and trimer phase is at $k_{s}=650$
approximately. (e) Stable values of the fractions $\phi_{i}$ as a
function of $k_{s}$.\label{fig:Model-and-passive}}
\end{figure*}

Here in the present work, we report theoretically an example that
active core corona particles can be successfully self-assembled into
a large scaled and highly ordered structure. In the absence of activity,
the extruding between core corona particles at dense systems strongly
suppress the local arrangements, thus the system would be trapped
into metastable states with much defects as mentioned above. We find,
however, if exerting an active force for each particle, the system
would form a highly ordered stripe pattern or trimer lattice\textcolor{black}{.
In particular, such an activity-induced ordered structure owns a self-healing
feature such that the system can maintain a very low defect ratio
as long as the activity exists. Besides, we also find that activity
has a two-fold effect: while a moderate activity can facilitate the
emergence of ordered structure, a strong activity could destroy the
order. Therefore, it is convenient to tune the system order by simply
changing the strength of activity.} Our results may predict a new
general routine to achieve high-ordered and self-healing structures
by simply introducing active force to self-assembly system.

\section{Result And Discussion}

We perform two-dimensional overdamped Langevin dynamics simulation
of 4096 active core corona particles (ACCP) in a periodic square box.
In Figure.\ref{fig:Model-and-passive}(a), a schematic diagram of
a pair of ACCPs is shown. The cores with diameter $\sigma$ model
the unpenetrable colloidal spheres described by Weeks-Chandler-Andersen
(WCA)-potential. The interaction between the ambient coronas, in consideration
of the essential entropy-elastic property of polymers or microgels\cite{cheng2010probing,rey2017anisotropic},
is described by a simple spring potential with equilibrium distance
$r_{c}=3\sigma$ and is also cutoff at $r_{c}$. The interaction as
a function of distance between two ACCPs is depicted by the curve
in Figure.\ref{fig:Model-and-passive}(a), which consists of a soft-repulsion
part in long-distance range between coronas and a strong-repulsion
part in short-distance range between cores. Additionally, each core
of ACCP is propelled by an active force $F_{a}$ (shown by the red
arrow in Figure.\ref{fig:Model-and-passive}(a)) with an orientation
undergoing random Brownian rotation. The square box length is set
to $100\sigma$, corresponding to a number density $c\simeq0.41$.
In the present work, the strength of $F_{a}$ and the strength of
spring potential $k_{s}$ are used as free parameters, and all parameters
are dimensionless. A more thorough description of simulation method
is provided in Supporting Information S1.

\subsection{Passive case}

We first consider the passive case with $F_{a}=0$. The strength of
spring potential $k_{s}$ is fixed to be 200 if not otherwise stated.
Figure.\ref{fig:Model-and-passive}(b) shows the stationary pattern
of the system after a long enough time, starting from a random initial
condition. Note that the coronas are not drawn out for simplicity.
Typically, there exist three kinds of local structure phases, i.e.,
stripe phase (green), trimer phase (yellow) and disorder phase (blue).
To identify which phase a particle belongs to, we can evaluate the
angle $\theta(0<\theta<180^{\circ})$ of it with its two nearest neighbors.
A particle is in the stripe phase if $\theta>160^{\circ}$, trimer
phase if $50^{\circ}<\theta<70^{\circ}$ and disorder phase otherwise.
As can be seen from the figure, a large portion of the pattern are
in ordered stripe and trimer phases. In Figure.\ref{fig:Model-and-passive}(c),
the fractions $\phi_{i}(i=S,\,T,\,D$ corresponding to stripe, trimer
and disorder, respectively) of each phase as functions of time $t$
are presented, starting from a disordered state with $\phi_{D}\simeq0.8$.
Clearly, the system relaxes very fast to the stationary state wherein
the values of $\phi_{i}$ reach stable values, say $\phi_{S}\simeq0.5,\,\phi_{T}\simeq0.1$
and $\phi_{D}\simeq0.4$. Such a mixed state with coexistence of stripe,
trimer and disorder phases is very stable (as can be seen from the
time dependencies of $\phi_{i}$) and we have not observed any obvious
changes to the configuration after very long simulation time. That
means thermal fluctuations in such a crowded system play an insignificant
role to the dynamic process, in consistent with previous experimental
observations\cite{osterman2007observation,rey2017anisotropic}.

One should note that the mixed state shown in Figure.\ref{fig:Model-and-passive}(b)
is actually a metastable state of the system, rather than the equilibrium
state with the lowest free energy. The equilibrium phase diagram in
terms of $k_{s}$ (see methods in ref\cite{zu2017forming,schoberth2016molecular,barkan2014controlled})
of the system is shown in Figure.\ref{fig:Model-and-passive}(d).
As one can see, the system shows two distinct equilibrium phases,
namely stripe and trimer, with the variation of control parameter
$k_{s}$ and the phase boundary is at $k_{s}\simeq650$. Therefore,
the most stable state of the system with lowest free energy for $k_{s}=200$
should be a ordered stripe phase rather than the mixed one in Figure.\ref{fig:Model-and-passive}(b).
The reason that one observes a mixed-phase state which is metastable
rather than the equilibrium stripe phase is that there might be a
rather high free energy barrier between them, and the attraction basin
of the metastable one is much larger than that of the equilibrium
one. Starting from a random initial condition, it is highly probable
that the system will get trapped into the local minimum of the free
energy landscape corresponding to the metastable state, and the high
barrier makes it very difficult for the system to jump into the equilibrium
state. It is also interesting to note that both the stripe and trimer
structures are anisotropic, while the interaction between the particles
are purely isotropic. Indeed, this reflects a kind of symmetry breaking.
In such a dense system, if particles are equally spaced from each
other, the system would obtain a high energetic cost since the corona
of each particle overlaps with all its nearest neighbors. If the system
chooses a stripe or trimer configuration, however, despite of close
packing with two nearest particles, a particle would keep away from
other ones as much as possible and finally minimize the energetic
cost\cite{malescio2004stripe}.

As indicated by Figure.\ref{fig:Model-and-passive}(d), the equilibrium
state of the system would change from stripe to trimer with increment
of $k_{s}$. For $k_{s}=200$ as shown in Figure.\ref{fig:Model-and-passive}(c),
the observed metastable state has more stripe particles than trimer
particles, $\phi_{S}>\phi_{T}$. In Figure.\ref{fig:Model-and-passive}(e),
we show how the stable values of the fractions $\phi_{i}$ change
with $k_{s}$. With increasing $k_{s}$, $\phi_{T}$ ($\phi_{S}$)
increases (decreases) monotonically as expected, while the fraction
of disordered phase does not change much. The observed patterns all
look similar to that shown in Figure.\ref{fig:Model-and-passive}(b),
except that the fractions of different phases are different. Therefore,
starting from a random initial condition, the system would finally
get trapped into a metastable mixed-phase state no matter what the
value $k_{s}$ is. Indeed, this makes it hard to obtain very ordered
structure in practical self-assembly processes and it is demanding
to find a way to overcome this difficulty.

\begin{figure*}
\begin{centering}
\includegraphics[width=1.7\columnwidth]{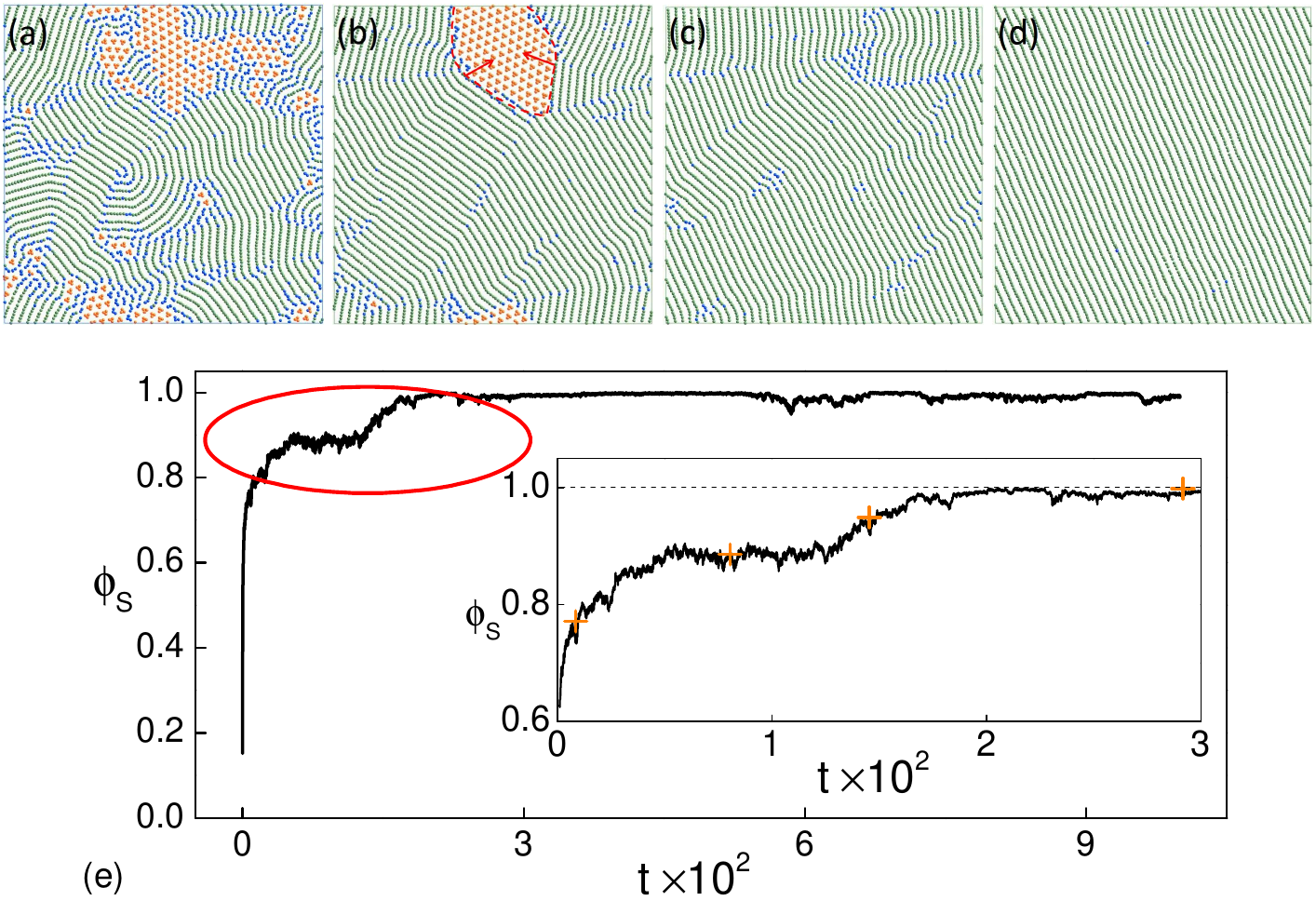}
\par\end{centering}

\caption{Formation of a highly stripe-ordered structure. (a)-(d) Representative
configurations at the stage $t=10$, $t=90$, $t=140$ and $t=290$,
respectively. The red dash line in (b) depicts a trimer domain and
two red arrows point the direction that trimer domain disappears.
(e) Fraction of stripe phase $\phi_{S}$ over time. Three dash lines
represents the change of $\phi_{S}$ of three independent systems,
and the solid line is a result over 20 times average. Inset is a zoom-in
of the area in red cycle, and four moments depicted by yellow crosses
correspond to (a)-(d) orderly. Note here $F_{a}=30$.\label{fig:stripe}}
\end{figure*}

\subsection{Active case}

We now put the system (still $k_{s}=200$) far from equilibrium by
adding a nonzero active force to each particle, starting from similar
random initial condition as for the passive case. In Figure.\ref{fig:stripe}(a)-(d),
we give typical snapshots of the system during the relaxation process
for an intermediate level of active force $F_{a}=30$. Very interestingly,
the final state shows a perfect stripe phase (Figure.\ref{fig:stripe}(d))
which is in sharp contrast to the metastable mixed-phase shown in
Figure.\ref{fig:Model-and-passive}(b) for $F_{a}=0$. This stripe
phase looks the same as the equilibrium one for $k_{s}=200$ shown
in Figure.\ref{fig:Model-and-passive}(d). Therefore, the system successfully
self-assemble into the demanded highly-ordered state by making the
particles active.

The transition from the initial disordered state (a) to the final
ordered state (d) consists of three main steps. In the first step
from (a) to (b), one can see that most of the disorder phase transform
into more ordered stripe or trimer phase, and the remaining trimer-phase
particles aggregate together to form a small island surrounded by
stripe phase. This trimer island will survive for some time, and since
stripe is more stable than trimer according to the equilibrium phase
diagram (Figure. \ref{fig:Model-and-passive}(d)) for $k_{s}=200$,
in the second step from (b) to (c), the trimer island will finally
be eliminated and the system changes into a state dominated overwhelmingly
by stripe phase. Such a stripe-dominated state still contains some
disorder defects, and in the third step from (c) to (d), the system
finally transforms into a nearly perfect stripe phase with no apparent
defects by local rearrangement and relaxation (see Movie 1). Accordingly,
the time dependence of $\phi_{S}$ is shown in Figure.\ref{fig:stripe}(e).
As can be seen, $\phi_{S}$ increases fastly at the very beginning
to a relatively high value about 0.78 (see the yellow cross in Figure.\ref{fig:stripe}(e)),
corresponding to the first step. Then $\phi_{S}$ keeps at a plateau
$\phi_{S}\simeq0.88$ for a quite long period of time. Clearly, the
plateau is related to the metastable trimer island which will survive
for some time before losing stability. The final slow increasing of
$\phi_{S}$ to 1.0 corresponds to the third relaxation step. Note
that the life time of the trimer island randomly changes from simulation
run to run, but the trimer island can always be observed during the
process.

\begin{figure*}
\begin{centering}
\includegraphics[width=1.7\columnwidth]{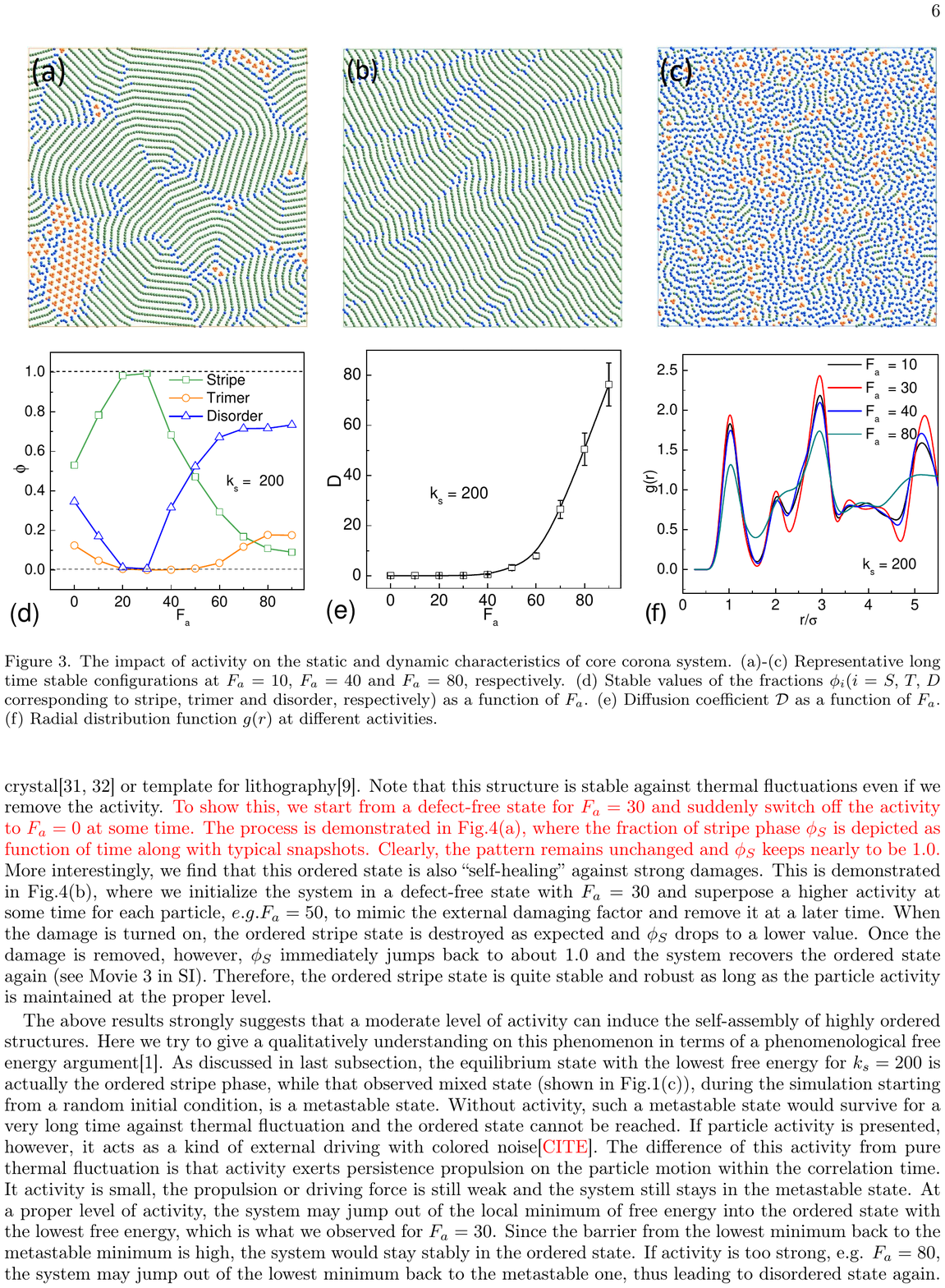}
\par\end{centering}

\caption{The impact of activity on the static and dynamic characteristics of
core corona system. (a)-(c) Representative long time stable configurations
at $F_{a}=10$, $F_{a}=40$ and $F_{a}=80$, respectively. (d) Stable
values of the fractions $\phi_{i}(i=S,\,T,\,D$ corresponding to stripe,
trimer and disorder, respectively) as a function of $F_{a}$. (e)
Diffusion coefficient $\mathcal{D}$ as a function of $F_{a}$. (f)
Radial distribution function $g(r)$ at different activities. \label{fig:The-impact-of}}
\end{figure*}

The above observations clearly demonstrate that particle activity
can induce very exciting behavior that is not available for the passive
counterpart system. Surely the results should be dependent on the
activity level $F_{a}$. In Figure.\ref{fig:The-impact-of}(a)-(c),
the final stationary states of the system for $F_{a}=10$, 40 and
80 are presented. In contrary to the case of $F_{a}=30$ shown in
Figure.\ref{fig:stripe}(d), these patterns are not fully ordered.
For a relatively small activity $F_{a}=10$, the trimer island can
not be eliminated by surrounding stripe phase such that the final
state is a mixture of stripe phase, trimer island and some disorder
defects. For $F_{a}=30$, as we already described in Figure.\ref{fig:stripe}(d),
a highly ordered stripe structure emerges. If increasing the activity
to $F_{a}=40$, however, the final state would not be that ordered,
as depicted in Figure.\ref{fig:The-impact-of}(b), wherein the system
maintains some long range order but with many dynamic defect boundaries
between ordered stripe clusters (see Movie 4). With further increasing
activity to a larger value, $\text{i.e.}$$F_{a}=80$, the system
turns into a totally disordered state as shown in Figure.\ref{fig:The-impact-of}(c).
Therefore, there exists an optimal level of activity (here $F_{a}\simeq30$)
that favorably supports the formation of ordered stripe phase. Accordingly,
the stable values of $\phi_{i}(i=S,\,T,\,D)$ are depicted in Figure.\ref{fig:The-impact-of}(d),
wherein $\phi_{S}$ non-monotonically changes with $F_{a}$ and reaches
the maximum about 1.0 for $F_{a}$ within a small range around 30.
Note that we have also studied a more larger system containing 16384
particles (see Supporting Information S2), and the same dynamic behaviors
are observed.

To get more information about the final stationary states of the system,
we have further investigated the dynamic and static characteristics
in terms of the diffusion coefficient $\mathcal{D}$ and the radial
distribution function $g(r)$, respectively. The diffusion coefficient
$\mathcal{D}$ is calculated via $\mathcal{D}=\ensuremath{\lim\limits _{t\to\infty}\frac{1}{4t}\langle\Delta r^{2}(t)\rangle}$,
where $\langle\Delta r^{2}(t)\rangle=\langle|\textbf{r}(t)-\textbf{r}(0)|^{2}\rangle$
is the mean square displacement (MSD) of a randomly chosen tagged
particle with $\textbf{r}(t)$ being the particle position at time
$t$. The radial distribution function $g(r)$, defined as $g(r)=\frac{1}{\rho^{2}}\langle\sum_{i=1}^{N}\sum_{j\neq i}^{N}\delta(\mathbf{r}-\mathbf{r}_{i})\delta(\mathbf{r}-\mathbf{r}_{j})\rangle$
where $\rho=N/V$ is the number density, characterizes the probability
of finding a pair of particles at a distance $r$ and thus reflects
the structure of the system. In Figure.\ref{fig:The-impact-of}(e),
$\mathcal{D}$ as a function of $F_{a}$ is shown. With increasing
$F_{a}$, $\mathcal{D}$ starts from nearly zero at the initial stage
and increases sharply at $F_{a}\simeq40$. This result indicates that
stable state of the system behaves like solids for small activity
$F_{a}\apprle40$. We note here the stable structure for $F_{a}=30$,
which is very ordered, is also very stable against thermal fluctuations
since the diffusion coefficient $\mathcal{D}$ is zero. If activity
is large, however, the system behaves more like fluids with large
$\mathcal{D}$ and loses long range order, as for example depicted
in Figure.\ref{fig:The-impact-of}(c) for $F_{a}=80$. The effect
of particle activity can also be reflected in $g\left(r\right)$ as
drawn in Figure.\ref{fig:The-impact-of}(f). Generally, $g\left(r\right)$
shows main peaks at $r/\sigma\simeq1$, 2.7 and 5.3, corresponding
to the typical nearest neighboring structures of the system. In particular,
the peak at $r/\sigma\simeq2.7$ corresponds to the distance between
neighboring parallel stripe segments, and that at $r/\sigma\simeq5.3$
to the distance between next-nearest parallel stripes. One can see
that the peaks at $2.7\sigma$ exist for all the activities shown,
suggesting that local stripe segments maintain with the variation
of $F_{a}$, and the distance between neighboring parallel segments
remains nearly unchanged as can be observed for all the patterns shown
in Figure.\ref{fig:The-impact-of}(a)-(c) and Figure.\ref{fig:stripe}(d).
For not strong activities $F_{a}=10$, 30 and 40, the peaks at $5.3\sigma$
are remarkable, indicating the existence of some long-range order
of parallel stripe patterns. The peak for $F_{a}=30$ is highest among
all them, corresponding to the most ordered stripe lattice given in
Figure.\ref{fig:stripe}(d). However, for a too strong activity, say
$F_{a}=80$, the peak at $5.3\sigma$ disappears, indicating that
the long-range stripe order is destroyed. All these findings clearly
suggest that activity can impose two-fold effects on this core-corona
system: an proper level of activity can facilitate the emergence of
ordered structure, while a large one may ruin the order and make the
system more disordered. Therefore, the system order shows a non-monotonic
dependency on particle activity, in consistent with Figure.\ref{fig:The-impact-of}(d).

\begin{figure}
\begin{centering}
\includegraphics[width=0.9\columnwidth]{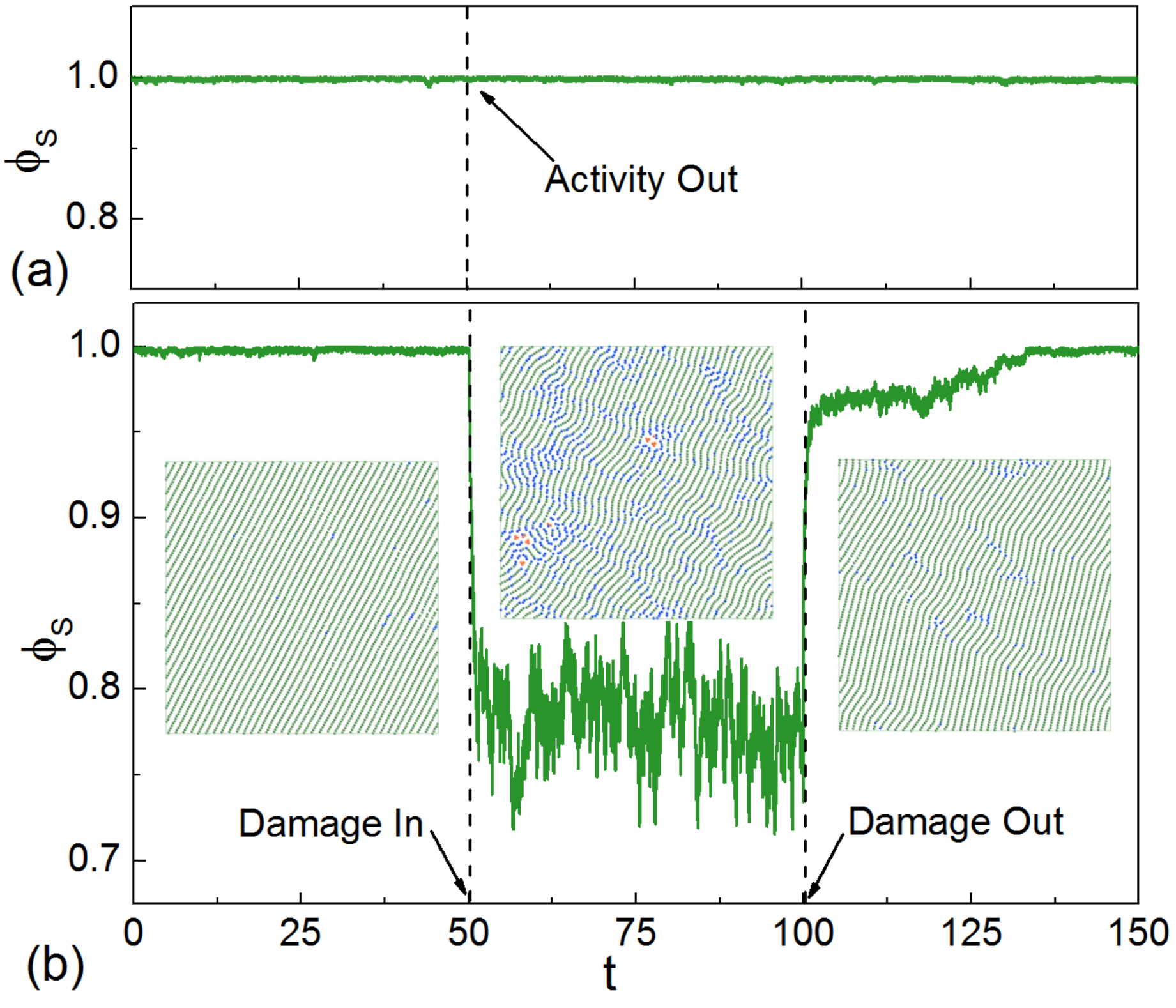}
\par\end{centering}

\caption{Stability of active induced ordered structure. (a) Against thermal
fluctuations. The system is initialized in an almost perfect stripe-ordered
structure ($\phi_{S}\simeq1.0$) with $k_{s}=200$ and $F_{a}=30$.
And the activity is switched to zero at $t=50$ as the black arrow
shown. (b) Against strong damages. The system is also initialized
in $\phi_{S}\simeq1.0$ with $k_{s}=200$ and $F_{a}=30$. At $t=50$,
an activity $F_{a}=50$ modeled as the damaging factor is superposed
and then it is removed at $t=100$ as the black arrows shown. \label{fig:Mechanism} }
\end{figure}

The emergence of highly ordered stripe structure at an optimal level
of activity (here $F_{a}=30$) demonstrated above, if achieved in
practice, could be very promising for material design for important
systems such as photonic crystal\cite{park2015structural,kolle2010mimicking}
or template for lithography\cite{rey2015fully}. Note that this structure
is stable against thermal fluctuations even if we remove the activity.
To show this, we start from a defect-free state for $F_{a}=30$ and
suddenly switch off the activity to $F_{a}=0$ at some time. The process
is demonstrated in Figure.\ref{fig:Mechanism}(a), where the fraction
of stripe phase $\phi_{S}$ is depicted as function of time. Clearly,
the pattern remains unchanged and $\phi_{S}$ keeps nearly to be 1.0.
More interestingly, we find that this ordered state is also ``self-healing''
against strong damages. This is demonstrated in Figure.\ref{fig:Mechanism}(b),
where we initialize the system in a defect-free state with $F_{a}=30$
and superpose a higher activity at some time for each particle, $e.g.$$F_{a}=50$,
to mimic the external damaging factor and remove it at a later time.
When the damage is turned on, the ordered stripe state is destroyed
as expected and $\phi_{S}$ drops to a lower value. Once the damage
is removed, however, $\phi_{S}$ immediately jumps back to about 1.0
and the system recovers the ordered state again (see Movie 3). Therefore,
the ordered stripe state is quite stable and robust as long as the
particle activity is maintained at the proper level.

The above results strongly suggests that a moderate level of activity
can induce the self-assembly of highly ordered structures. Here we
try to give a qualitatively understanding on this phenomenon in terms
of a phenomenological free energy argument\cite{boles2016self}. As
discussed in last subsection, the equilibrium state with the lowest
free energy for $k_{s}=200$ is actually the ordered stripe phase,
while that observed mixed state (shown in Figure.\ref{fig:Model-and-passive}(b)),
with the simulation starting from a random initial condition, is a
metastable state. Without activity, such a metastable state would
survive for a very long time against thermal fluctuation and the ordered
state cannot be reached. If particle activity is presented, however,
it acts as a kind of external driving with colored noise\cite{fodor2016far}.
The difference of this activity from pure thermal fluctuation is that
activity exerts persistence propulsion on the particle motion within
the correlation time. It activity is small, the propulsion or driving
force is still weak and the system still stays in the metastable state.
At a proper level of activity, the system may jump out of the local
minimum of free energy into the ordered state with the lowest free
energy, which is what we observed for $F_{a}=30$. Since the barrier
from the lowest minimum back to the metastable minimum is high, the
system would stay stably in the ordered state. If activity is too
strong, e.g. $F_{a}=80$, the system may jump out of the lowest minimum
back to the metastable one, thus leading to disordered state again.
Surely such a phenomenological description is quite qualitative and
a theoretical analysis would be highly helpful, however, it is beyond
the scope of current study and may deserve a separate work.

\begin{figure}
\begin{centering}
\includegraphics[width=0.9\columnwidth]{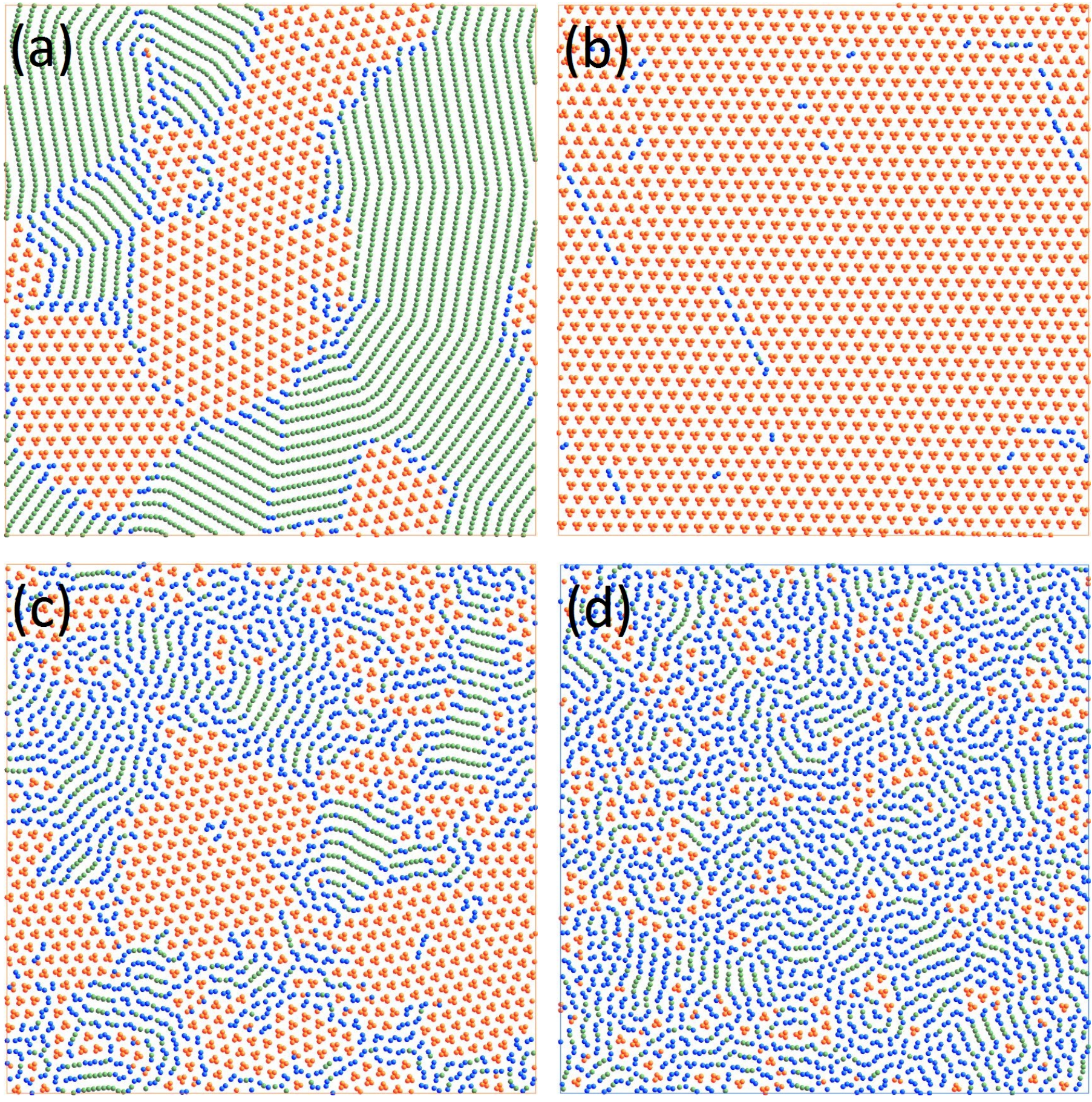}
\par\end{centering}

\caption{Representative long time stable configurations at $k_{s}=300$ for
$F_{a}=20$, 60, 80 and 120, corresponding to (a)-(d), respectively.\label{fig:Trimer-case} }
\end{figure}

So far, we have studied the results for $k_{s}=200$, where the equilibrium
state is the stripe phase. It is then interesting to investigate the
phase behaviors for other values of $k_{s}$. In Figure.\ref{fig:Trimer-case}(a)-(d),
four representative stable patterns for $k_{s}=300$ with $F_{a}=20,$
60, 80 and 120 are shown, respectively. For a low activity $F_{a}=20$
in (a) , the system is now typically a mixture of trimer phase and
stripe phase. For a moderate activity $F_{a}=60$ in (b), interestingly,
a highly ordered trimer phase appears. The formation of such an ordered
trimer phase is similar to that of the stripe phase shown in Figure.\ref{fig:stripe},
involving a relatively fast process forming several big clusters of
trimer phases and a slow relaxation process healing the grain boundaries
(see Movie 2). However, there are still many defects doped linearly
inside the lattice, which is slightly different to a $k_{s}=200$
case that the system contains nearly no defects. This is due to that
these defects are usually trapped as dimmers, and they could hardly
be merged into trimers only when at least 3 of them meet. Note that
this ordered structure also has a self-healing feature (see Supporting
Information S3). Further increasing $F_{a}$ to 80 in (c), such a
trimer superlattice becomes unstable and the system contains a few
clusters of trimer surrounded by disordered phase or small stripe
phase. Compared to the state shown in (a), the system now is more
``living''\cite{nguyen2014emergent}, i.e, the clusters may split
into small pieces and then merge again (see Movie 5) while that in
(a) is static. If the activity is even higher, say for $F_{a}=120$
in (d), the system behaves like liquids and becomes disorder. The
non-monotonic dependency between activity $F_{a}$ and the fraction
of trimer phase $\phi_{T}$ (see Supporting Information S4) is very
similar to the $k_{s}=200$ case.

\begin{figure*}
\begin{centering}
\includegraphics[width=1.7\columnwidth]{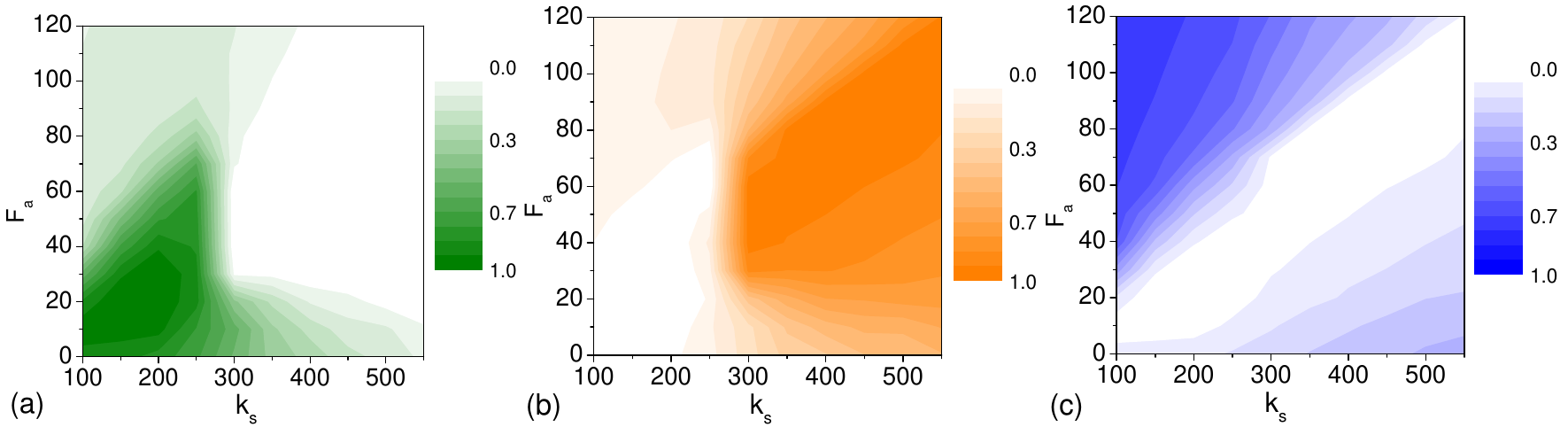}
\par\end{centering}

\caption{Non-equilibrium phase diagram of long time stable value of (a)$\phi_{S}$
colored in green, (b)$\phi_{T}$colored in yellow and (c)$\phi_{D}$
colored in blue, respectively. The color bar attached at right-side
of each phase diagram is at a range of (0, 1.0). \label{fig:phase-diagram.}}
\end{figure*}

It is interesting to note that the activity-induced ordered structure
for $k_{s}=300$ and $F_{a}=60$ is the trimer phase rather than the
stripe phase, while the equilibrium state for $k_{s}=300$ is still
stripe phase as suggested by Figure.\ref{fig:Model-and-passive}(d).
This indicates that activity can not only facilitate the self-assembly
process to reach the ordered structure, but also can induce new structure
that is not available in the equilibrium counterpart system. It is
then very demanded to obtain a global picture by investigating the
whole (nonequilibrium) phase behavior in the parameter plane spanned
by $F_{a}$ and $k_{s}$. In Figure.\ref{fig:phase-diagram.}(a)-(c),
the contour plots of $\phi_{S}$, $\phi_{T}$ and $\phi_{D}$ in $(F_{a},\,k_{s})$
plane are presented, respectively. In (a), one can see that $\phi_{S}$
is highest in the region of small $k_{s}$ and $F_{a}$ and disappears
if $k_{s}$ exceeds some threshold value. In (b), the distribution
of $\phi_{T}$ is just complementary to that of $\phi_{S}$, i.e.,
it is highest in the region where $k_{s}$ and $F_{a}$ are large.
The transition from stripe to trimer approximately takes place at
$k_{s}\simeq250$ , which is not sensitive to the value of $F_{a}$
in the moderate range. Note that the transition from stripe to trimer
for an equilibrium system takes place at nearly $k_{s}\simeq650$,
suggesting that activity strongly shifts the transition point. Therefore,
in a broad range of $k_{s}\in(250,\,650)$, activity can induce a
transformation from stripe to trimer. For a given $k_{s}$, both $\phi_{S}$
and $\phi_{T}$ show non-monotonic dependencies on $F_{a}$, as can
be seen from these two figures. In (c), it is shown that the system
is quite disordered in the left-up and right-bottom corners of the
parameter plane. In the diagonal region, $\phi_{D}$ is zero, indicating
that the system is in purely ordered state without defects. The optimal
value of $F_{a}$ seems increases linearly with $k_{s}$ as suggested
by the contour plot.

\section{Conclusion }

In summary, we have theoretically studied the self-assembly process
of active core corona particles by using two-dimensional overdamped
Langevin dynamics simulation. This system with typically two-length
scales owns the possibility to form various anisotropic and super
lattice structures. However, such a system would generally be trapped
into a metastable state with many defects and the transition from
metastable state to a thermodynamically favored one involves many
local structural rearrangements, which makes the emergence of target
ordered structure occur at a unreachable long time scale. However,
if exerting a moderate active force to each particle, such a metastable
state can successfully cross the energy barrier leading to a large-scaled
and highly ordered stripe or trimer lattice. In addition, these active
induced ordered structures are stable against thermal fluctuations
even when the activity is removed and can spontaneously heal strong
damages as long as the activity keeps at a proper level. Our results
demonstrate clearly that activity can be used as a new design strategy
for self-assembly process.

It would be interesting to test our model in experimental systems,
for instance, dense system of Janus particles coated by microgels
in alternating electric field. The particles could exhibit spontaneous
motility in such field\cite{yan2016reconfiguring,bricard2013emergence}
and repel each other due to the ambient soft microgels\cite{rey2017anisotropic}.
In practice, spring potential may not exactly hold when two coronas
are strongly extrude, while one can alternatively use an external
field to achieve a more robust repulsive interaction. For example,
N. Osterman $et.al$ used an external magnetic field to achieve the
repulsion between superparamagnetic spheres which are trapped in a
thin wedge-shaped cell, and softened the repulsion by changing the
cell thickness\cite{osterman2007observation}. Since the two key ingredients,
corona\cite{dotera2014mosaic,zu2017forming,rey2017anisotropic} and
activity\cite{patteson2018propagation,geyer2018sounds,wu2017transition,mallory2019activity,mallory2018active},
are ubiquitous in many real systems and widely studied both theoretically
and experimentally, we hope that our results could open new perspectives
to the fabrication of similar highly ordered and self-healing structure
and new insights to defect engineering in material field.
\begin{acknowledgments}
This work is supported by MOST(2016YFA0400904, 2018YFA0208702), by
NSFC (21833007, 21790350, 21673212, 21521001, 21473165, 21403204),
by the Fundamental Research Funds for the Central Universities (2340000074),
and Anhui Initiative in Quantum Information Technologies (AHY090200).
\end{acknowledgments}

%\bibliographystyle{apsrev4-1}
%\addcontentsline{toc}{section}{\refname}\bibliography{DRC,Bib_core_corona,Bib_active,ACCP}
%

\end{document}